\documentclass[prb, twocolumn, showpacs]{revtex4}

\usepackage{amsfonts}
\usepackage{amsmath}
\usepackage{subfigure}
\usepackage{tikz}

\newcommand{\ca}[1]{{\cal #1}}
\newcommand{\ff}[1]{{\boldsymbol #1}}

\begin{document}

\author{Philipp Jurgenowski}
\author{Michael Potthoff}

\affiliation{I. Institut f\"ur Theoretische Physik, Universit\"at Hamburg, Jungiusstra\ss{}e 9, 20355 Hamburg, Germany}



\title{Dynamical symmetry between spin and charge excitations studied by a plaquette mean-field approach in two dimensions}

\begin{abstract}
The real-time dynamics of local occupation numbers in a Hubbard model on a $6\times 6$ square lattice is studied by means of the non-equilibrium generalization of the cluster-perturbation theory.
The cluster approach is adapted to studies of two-dimensional lattice systems by using concepts of multiple-scattering theory and a component decomposition of the non-equilibrium Green's function on the Keldysh-Matsubara contour.
We consider ``classical'' initial states formed as tensor products of states on $2\times 2$-plaquettes and trace the effects of the inter-plaquette hopping in the final-state dynamics. 
Two different initially excited states are considered on an individual plaquette, a fully polarized staggered spin state (N\'eel) and a fully polarized charge-density wave (CDW). 
The final-state dynamics is constrained by a dynamical symmetry, i.e.\ the time-evolution operator and certain observables are invariant under an anti-unitary transformation composed of time reversal, an asymmetric particle-hole, and a staggered sign transformation.
We find an interesting interrelation between this dynamical symmetry and the separation of energy and time scales: 
In case of a global excitation with all plaquettes excited, the initial N\'eel and the initial CDW states are linked by the transformation. 
This prevents an efficient relaxation of the CDW state on the short time scale governing the dynamics of charge degrees of freedom.
Contrary, the CDW state is found to relax much faster than the N\'eel state in case of a local excitation on a single plaquette where the symmetry relation between the two states is broken by the coupling to the environment.
\end{abstract}

\maketitle

\section{Introduction}

Due to recent substantial progress in the field of ultracold quantum gases in optical lattices,\cite{Jaksch_and_Zoller_2005,Lewenstein_et_al_2007,Bloch_et_al_2008,Hung_et_al_2010,Strohmaier_et_al_2010}
an almost complete real-time control of the parameters of a quantum system has become experimentally feasible.
The study of ultracold fermions in optical lattices thus offers unique new possibilities to understand the non-equilibrium dynamics of quantum systems.
Symmetries play an important role for the quantum dynamics.
Particularly interesting are ``dynamical symmetries'' where the time-dependent expectation value of an observable is the same for certain different initial states due to the invariance of the observable as well as of the time-evolution operator under an (anti-)unitary transformation (see Appendix \ref{sec:dsym}).
Recently, the expansion of an initially confined quantum gas of fermions in the lowest band of a homogeneous optical lattice has been studied after suddenly switching off the confining potential, and the observed independence of the dynamics on the sign of the (Hubbard-type) interaction could be explained by a dynamical symmetry.\cite{Schneider_et_al_2012}

Here, we study a different situation where a dynamical symmetry holds {\em locally} for two different initial states but is broken due to their coupling to the environment.
We consider the dynamics of a system of strongly interacting spin-1/2 fermions in a two-dimensional Hubbard model at half-filling on a square lattice with nearest-neighbor hopping.
This has become accessible to studies of ultracold quantum gases.\cite{Koehl_et_al_2005,Joerdens_et_al_2008}
The local density of spin-$\uparrow$ particles, $n_{j\uparrow}$, is our observable of interest.
It has been demonstrated recently with bosons that the parity of the number of particles can be measured with single-site resolution even in a strongly correlated system. \cite{Bakr_et_al_2009,Bakr_et_al_2010,Sherson_et_al_2010}
For the Hubbard model and observables $n_{j\uparrow}$, we will argue that a particular combination $\ca U$ of time-reversal, an asymmetric particle-hole and a staggered sign transformation represents a dynamical symmetry. 

An interesting question is how fast an initially prepared ``classical'' state builds up entanglement and correlations.
This can be studied by preparing the initial state as the ground state of the model with isolated plaquettes $p$, i.e.\
$| \Psi(t_{0}) \rangle = \otimes_{p}^{{\rm plaquettes}} | p (t_{0}) \rangle$, while for the subsequent time evolution ($t>t_{0}$) the inter-plaquette hopping $\ff V$ is suddenly switched on at $t=t_{0}$.
Double-well systems in one dimension\cite{Sebby-Strabley_et_al_2006,Anderlini_et_al_2007,Foelling_et_al_2007,Trotzky_et_al_2008} as well as plaquette systems in two dimensions\cite{Nascimbene_et_al_2012} have recently been studied experimentally.

A non-trivial dynamics, already for an isolated pla\-quet\-te, is obtained by preparing the plaquette initial state with the help of local fields producing a classical N\'eel state with staggered spins $| p(t_{0}) \rangle = | \text{N\'eel}, p \rangle$ or a charge-density wave $| p(t_{0}) \rangle = | \text{CDW}, p \rangle$ and by suddenly switching off the field at $t=t_{0}$.
In case of full polarization, 
these states transform into each other under $\ca U$ while the plaquette ground state in the absence of the fields, $|p(t_{0}) \rangle = | {\rm GS}, p \rangle$, is not invariant under $\ca U$.
By comparing the time evolution of $n_{j\uparrow}$ starting from the initial plaquette product state with a global N\'eel excitation with the corresponding one for a global CDW excitation, we study the effects of the dynamical symmetry. 
In case of a N\'eel or CDW excitation that is localized to a single plaquette, we expect a breakdown of the dynamical symmetry due to the inter-plaquette coupling.

Another goal of the present paper is a methodical one:
Studying the real-time dynamics of a strongly-correlated lattice-fermion model in two dimensions far from equilibrium is a highly non-trivial task. 
Standard methods as dynamical density-matrix renormalization group\cite{Schollwoeck_2005} or quantum Monte-Carlo\cite{Gull_et_al_2011} approaches cannot be used in this case. 
We therefore resort to a cluster (plaquette) mean-field approach that has been developed recently for one-dimensional and impurity-type models, namely the non-equilibrium variant \cite{Balzer_and_Potthoff_2011,KvdLA11,Balzer_et_al_2012,Nuss_et_al_2012} 
of the cluster-perturbation theory (CPT).\cite{Gros_and_Valenti_1993,Senechal_et_al_2000,Senechal_et_al_2002,Zacher_et_al_2002}
The NE-CPT can be characterized as a time-dependent multiple-scattering approach. 
It is exact for the non-interacting model with Hubbard-$U=0$ and in the case of disconnected clusters, i.e.\ $\ff V=0$. 
In the general case, the NE-CPT starts from the one-particle propagator of the system of isolated clusters as a reference point and includes the effects of scattering at the inter-cluster potential $\ff V$ perturbatively to all orders but neglecting certain vertex corrections.
Up to now, to study transient dynamics, the NE-CPT has been applied to one-dimensional or impurity-type models only.
Here, our goal is to extend the theory and its implementation to two-dimensional models which are not easily accessible by other techniques.

\section{Nonequilibrium cluster-perturbation theory}

To discuss the main idea of the non-equilibrium cluster-perturbation theory (NE-CPT) and the necessary steps for an extension to two-dimensional lattice fermion models, we consider the single-band Hubbard model.\cite{Hubbard} 
Using standard notations, the Hamiltonian that governs the time evolution of the system for $t>t_{0}$ is given by:
\begin{eqnarray}
H(t)
&=& 
\sum_{j,k,\sigma} T^{H}_{jk}(t) c_{j\sigma}^{\dagger} c_{k\sigma}
+ 
U^{H}(t)
\sum_{j} n_{j\uparrow} n_{j\downarrow} 
\nonumber \\
&+& 
\sum_{j,k,\sigma} V^{H}_{jk}(t) \, c_{j\sigma}^{\dagger} c_{k\sigma} 
\: .
\label{eqn:hamh}
\end{eqnarray}
We formally allow for an explicit time dependence of $H(t)$ but will restrict ourselves to $H=\mbox{const.}$ for the evaluation of the theory. 
In the latter case, the time evolution operator is given by $\exp(-i  H  (t-t_{0}))$. 

The whole lattice is split into disconnected finite ``clusters''.
The first two terms on the r.h.s.\ of Eq.\ (\ref{eqn:hamh}) represent the intra-cluster Hamiltonian with the intra-cluster hopping $\ff T^{H}$ and the on-site Hubbard interaction $U^{H}$. 
Note that the locality of the interaction part is a necessary ingredient for the CPT.
The third term represents the hopping interlinking the different clusters, and the corresponding matrix elements are denoted by $\ff V^{H}$.
The Hilbert space associated with an isolated cluster is assumed to be sufficiently small such that all cluster physical quantities of interest can be calculated exactly by numerical means. 
For the subsequent calculations we address a two-dimensional square lattice which is tiled into quadratic plaquettes consisting of four sites only. 

The time evolution is assumed to start at $t=t_{0}$ from a pure initial state $| \Psi(t_{0}) \rangle$ which is taken to be the ground state of a Hamiltonian $\ca B$. \cite{Wagner_1991}
This has basically the same structure as $H(t)$ but with possibly different parameters:
\begin{eqnarray}
\ca B 
& \equiv &
B - \mu N 
= 
\sum_{j,k,\sigma} \left(T^{B}_{jk} - \mu \delta_{jk}\right) c_{j\sigma}^{\dagger} c_{k\sigma} 
\nonumber \\
&+& 
U^{B} \sum_{j} n_{j\uparrow} n_{j\downarrow} 
+ 
\sum_{j,k,\sigma} V^{B}_{jk} \, c_{j\sigma}^{\dagger} c_{k\sigma}
\: .
\label{eqn:hamb}
\end{eqnarray}
The chemical potential $\mu$ can be adjusted to get the desired total particle number in the initial ground state.
As the ground state of a correlated many-body system, the initial state must be treated by approximate means. 
Hence, to prepare for the CPT, we consider the same tiling of the lattice and the same corresponding splitting of the Hamiltonian.
Note that a generalization to finite temperatures and thus a mixed, e.g.\ grand-canonical, equilibrium initial state could easily be described but will not be considered here.

\begin{figure}[t]
\includegraphics[width=0.36\textwidth]{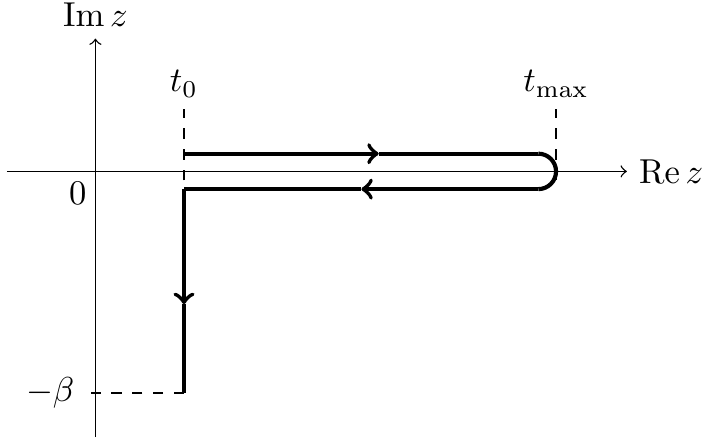}
\caption{\label{fig:contour} 
Contour $\mathcal{C}$ in the complex time plane.
$t_{0}$ is the time at which the system is prepared in the initial state.
$t_{\text{max}}$ is the maximum time up to which observables are traced.
$\beta$ is the inverse temperature.
For a pure initial state $| \Psi(t_{0}) \rangle$ we have $\beta \rightarrow \infty$.
}
\end{figure}

The time evolution of our basic quantity of interest, namely the expectation value $\langle \widehat{n}_{j\uparrow} (t) \rangle$ of the local spin-$\uparrow$ occupation number at site $j$, but actually the time evolution of all one-particle observables $A(t) = \sum_{j,k,\sigma} a_{jk\sigma}(t) c^{\dagger}_{j\sigma} c_{k\sigma}$ with possibly explicitly time-dependent parameters, can be obtained from the non-equilibrium one-particle Green's function
\begin{equation}
\label{eqn:gfdef}
G^{\sigma}_{jk}(z_{1},z_{2}) \equiv -i \, \Bigl\langle \mathcal{T}_{\mathcal{C}} \, \Bigl( \widehat{c}_{j\sigma}(z_{1}) \widehat{c}^{\dagger}_{k\sigma}(z_{2}) \Bigr) \Bigr\rangle
\end{equation}
via
\begin{equation}
\label{eqn:gf-obs}
\bigl\langle \widehat{A}(t) \bigr\rangle = -i\sum\limits_{j,k,\sigma} a_{jk\sigma}(t) \, G^{\sigma}_{kj} (t^{+},t^{-}) \; .
\end{equation}
Here, $z_{1}$ and $z_{2}$ run over the Keldysh-Matsubara contour $\mathcal{C}$ in the complex time plane \cite{Danielewicz_1984,Wagner_1991} which is shown in Fig.\ \ref{fig:contour}.
It consists of three branches: 
the upper and the lower Keldysh branch along the real axis ($\mathcal{K}^{+}$ and $\mathcal{K}^{-}$, respectively) as well as the Matsubara branch $\mathcal{M}$ parallel to the imaginary one. 
Here, $t$ stands for physical (real) times while $t_{0} - i\tau$ denotes a time point on $\mathcal{M}$, and $t^{\pm} \in \mathcal{K}^{\pm}$. 
Furthermore, $\mathcal{T}_{\mathcal{C}}$ is the time-ordering operator on $\mathcal{C}$. 
Operators with a hat are given in their Heisenberg picture.
Expectation values as well as Heisenberg time evolution refer to ``fully interacting'' Hamiltonians $\mathcal{B}$ and $H(t)$, respectively.

As opposed to a direct calculation, the main advantage to compute one-particle quantities (Eq.\ (\ref{eqn:gf-obs})) from the double-time propagator $\ff G$ (Eq.\ (\ref{eqn:gfdef})) on the Keldysh-Matsubara contour is that this is accessible to standard perturbative techniques, see Ref.\ \onlinecite{Wagner_1991}.
Here, we expand the full $\ff G$ in powers of the inter-plaquette hopping $\ff V$.
The ($\ff V=0$)-propagator $\ff G'$ must be computed for the fully interacting model but for isolated plaquettes with small Hilbert space only:
${G'}^{\sigma}_{jk}(z_{1},z_{2}) \equiv 0$ for lattice sites $j$ and $k$ belonging to different plaquettes.
An according numerically exact Krylov-space technique for an efficient computation of $\ff G'$ has been introduced in Ref.\ \onlinecite{Balzer_et_al_2012}. 

Starting from $\ff G'$, the propagator of the whole system can be obtained by summing inter-plaquette-scattering diagrams up to infinite order in $\ff V$:
\begin{equation}
\label{eqn:cpt-prop}
\ff G^{\text{(CPT)}} = \ff G' + \ff G' \bullet \ff V \bullet \ff G' + \cdots \: .
\end{equation}
For $U^{B}=U^{H}(t)=0$ this procedure represents an exact time-dependent multiple-scattering approach as all scattering paths are summed over. 
Furthermore, it is trivially exact in the case $\ff V=0$.
For finite interactions and finite inter-cluster hopping, however, the NE-CPT propagator Eq.\ (\ref{eqn:cpt-prop}) is approximate as certain vertex corrections are neglected (see discussion in Ref.\ \onlinecite{Balzer_and_Potthoff_2011}).
Alternatively, the NE-CPT may be seen as an exact approach on the length scale of an individual cluster but as mean-field-like beyond.

A main disadvantage of the approach consists in the lack of any self-consistency.
This is opposed to more elaborate cluster mean-field approaches \cite{Potthoff_2012} as the cellular dynamical mean-field theory \cite{Kotliar_et_al_2001} or the dynamical cluster approximation \cite{Hettler_et_al_1998,Hettler_et_al_2000} which, on the other hand, are not easily applicable to the non-equilibrium case. 
Lack of self-consistency or lack of variational optimization also implies that the approach cannot be expected to respect macroscopic conservation laws, i.e.\ conservation of the total particle number and the total spin, that result from the U(1) and the SU(2) symmetry of the Hamiltonian.\cite{Baym_and_Kadanoff_1961,Baym_1962}
We therefore restrict the application of the method to the case of the Hubbard model at half-filling and vanishing $z$-component of the total spin, i.e.\ $L_{p}^{-1} \sum_{j\in p} \langle \widehat{n}_{j\sigma} (t) \rangle = 0.5$ for each plaquette $p$ with $L_{p}=4$ sites. 
In this case particle-number and spin conservation is enforced by manifest particle-hole and spin-inversion symmetry. 

A re-summation of the terms on the r.h.s.\ of Eq.\ (\ref{eqn:cpt-prop}) yields the NE-CPT equation
\begin{equation}
\label{eqn:cpt-eqn}
\ff G^{\rm (CPT)} = \ff G' + \ff G' \bullet \ff V \bullet \ff G^{\rm (CPT)}
\end{equation}
which represents an obvious generalization of the static CPT\cite{Gros_and_Valenti_1993,Senechal_et_al_2000,Senechal_et_al_2002,Zacher_et_al_2002} to the time-dependent case.
Here, all quantities are to be understood as having one spin index, two spatial indices as well as two time arguments.
Furthermore, $\ff C = \ff A \bullet \ff B$ is short for $C^{\sigma}_{jk} (z_{1},z_{2}) = \sum_{l} \int_{\mathcal{C}}\!dz_{3} \, A^{\sigma}_{jl} (z_{1},z_{3}) B^{\sigma}_{lk} (z_{3},z_{2})$. 
For the inter-cluster hopping we have
$V^{\sigma}_{jk}(z_{1},z_{2}) = V_{jk}(z_{1}) \delta_{\mathcal{C}}(z_{1},z_{2})$ where $\delta_{\mathcal{C}}$ is the contour delta function, and $V_{jk}(z) = V^{H}_{jk}(t)$ for $z \in \mathcal{K}^{\pm}$ while $V_{jk}(z) = V^{B}_{jk}$ for $z \in \mathcal{M}$. 
Note that there are two independent CPT equations for each of the two spin species. 

\section{Iterative coupling of plaquettes}

In previous studies of transient dynamics,\cite{Balzer_and_Potthoff_2011,Balzer_et_al_2012} the NE-CPT has been applied to impurity-type models or small one-dimensional systems only. 
Furthermore, studies have so far been restricted to problems where the CPT equation (Eq.\ (\ref{eqn:cpt-eqn})) had to be solved only {\em once} for each parameter set.
For the intended application to a two-dimensional lattice, however, several plaquettes must be coupled, see Fig.\ \ref{fig:newsystem} for an example of an $8 \times 8$ square lattice with open boundaries.
In principle this could be achieved via Eq.\ (\ref{eqn:cpt-eqn}) in a {\em single} step by treating the inter-plaquette hopping between all plaquettes as the perturbation $\ff V$ {\em simultaneously}.
Alternatively, however, the Green's function of the entire lattice may be obtained in several basic steps where in each step only two (possibly  different) clusters (consisting of one or more plaquettes) are coupled at a time.
This second procedure just corresponds to an iteration of the elementary CPT concept and also works for the non-equilibrium case. 
It is easily seen to be equivalent with the standard (NE-)CPT. 
However, it is clearly more flexible conceptually as one can make use of known concepts of standard multiple-scattering theory: 

\begin{figure}[t]
\includegraphics[width=0.3\textwidth]{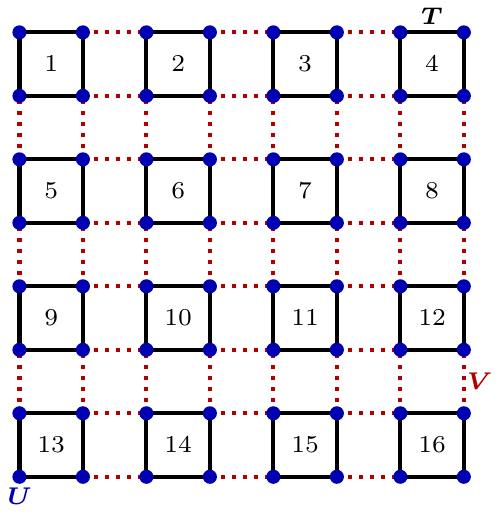}
\caption{\label{fig:newsystem} (Color online)
Decomposition of the Hubbard model on the two-dimensional square lattice into disconnected plaquettes. 
$T$ denotes the intra- and $V$ the inter-plaquette hopping. 
$U$ is the on-site Hubbard interaction. 
See text for discussion.
}
\end{figure}

Consider the case displayed in Fig.\ \ref{fig:newsystem} as an example and assume that $U^{H}(t) = U^{B} =\mbox{const.}$ and $\ff V^{H}(t) = \ff V^{B} = \mbox{const.}$ while $T_{jk}^{H}(t) = T_{jk}^{B} = \mbox{const.}$ for all sites $j,k$ {\em except for} the sites belonging to the plaquette $p=1$ in the upper left corner. 
For this example, it is clearly beneficial to couple the cluster ``1'' consisting of plaquette $p=1$ only, which is described by an intra-cluster Green's function $\ca G'_{1}$, with cluster ``2'' consisting of plaquettes $p=2$, ..., $p=16$, which is described by the intra-cluster Green's function $\ca G'_{2}$.
Namely, $\ca G'_{1}$ is easily obtained as the corresponding Hilbert space is small while the computation of $\ca G'_{2}$ is strongly simplified as this is an equilibrium problem. 
$\ca G'_{2}$ can thus be obtained by equilibrium CPT, coupling plaquettes $p=2$, ..., $p=16$ step by step. 

As a second example let us again consider the system displayed in Fig.\ \ref{fig:newsystem} but now we assume that the (time-dependent) inter-cluster hopping between all plaquettes is the same and that all plaquettes have identical (time-dependent) intra-cluster parameters.
In this case, a ``cluster-doubling method'' is helpful:
In a first step plaquettes $p=1$ and $p=2$ are coupled by means of the NE-CPT. 
The Green's function of the resulting cluster ``1+2'' is the same as the one of cluster ``3+4''.
Hence the latter is trivially obtained as a copy. 
In a second step, the clusters ``1+2'' and ``3+4'' are coupled to ``1+2+3+4'' which is again equivalent with ``5+6+7+8''.
This method can be iterated and is highly efficient.

We may therefore concentrate on the basic step of coupling two (possibly different) clusters:
Arranging the site indices in two blocks referring to the two clusters, we have 
\begin{equation}
\ff G' = \begin{pmatrix} \ca G'_{1} & 0 \\ 0 & \ca G'_{2} \end{pmatrix} \,, \qquad 
\ff V = \begin{pmatrix} 0 & \mathcal{V}\\ \mathcal{V}^{\dagger} & 0 \end{pmatrix} 
\end{equation}
for the $\ff V=0$ propagator and for the perturbation $\ff V$ itself while the CPT Green's function reads
\begin{equation}
 \ff G^{\rm (CPT)} = \begin{pmatrix} \ca G_{11} & \mathcal{G}_{12}\\ \mathcal{G}_{21} & \mathcal{G}_{22} \end{pmatrix} \,.
\end{equation}
The CPT equation (Eq.\ (\ref{eqn:cpt-eqn})) acquires a $2\times 2$ block structure and can be rewritten in the following way:
\begin{eqnarray}
\nonumber
\mathcal{G}_{11} 
& = & 
{\mathcal{G}}'_{1} + {\mathcal{G}}'_{1} \bullet \mathcal{V} \bullet {\mathcal{G}}'_{2} \bullet \mathcal{V}^{\dagger} \bullet \mathcal{G}_{11} \,,\\
\nonumber
\mathcal{G}_{22} 
& = & 
{\mathcal{G}}'_{2} + {\mathcal{G}}'_{2} \bullet \mathcal{V}^{\dagger} \bullet {\mathcal{G}}'_{1} \bullet \mathcal{V} \bullet \mathcal{G}_{22} \,,\\
\nonumber
\mathcal{G}_{12} 
&=& 
{\mathcal{G}}'_{1} \bullet \mathcal{V} \bullet \mathcal{G}_{22} \,,\\
\label{eqn:cptblock} 
\mathcal{G}_{21} 
&=& 
{\mathcal{G}}'_{2} \bullet \mathcal{V}^{\dagger} \bullet \mathcal{G}_{11} \,.
\end{eqnarray}
After solving the former two equations for the diagonal elements of $\ff G^{\text{(CPT)}}$, the non-diagonal ones can be easily obtained by evaluating the latter two. 
For a one-dimensional tight-binding system the matrix $\mathcal{V}$ exhibits exactly one non-zero element, and the sums over spatial indices, implicit in the $\bullet$~operation, reduce to just a single term.
In higher dimensions typically more than a single inter-cluster hopping term connect the same two clusters. 
Still the CPT equations represent a strongly sparse linear system of equations. 
Exploiting this sparsity is inevitable for an efficient implementation of the theory.

\section{Model and numerical approach}

As an example for the application of the non-equilibrium CPT to a two-dimensional system and for a discussion of dynamical symmetries (see Appendix \ref{sec:dsym}), we consider the two-dimensional Hubbard model on a square lattice at half-filling with hopping $T^{{B}}_{jk}=T^{{H}}_{jk}=T$ between nearest neighbors $j,k$ only and with $U^{B}=U^{H}=U$.
The energy and the time scales are set by $T=-1$.
The time-evolution operator is invariant (see Eq.\ (\ref{eqn:timeevo})) under an anti-unitary transformation
\begin{equation}
\ca U = \ca U_{1}  \ca U_{2}  \ca U_{3} \: .
\end{equation}
Here, $\ca U_{1}$ is the particle-hole transformation for the spin-$\downarrow$ particles,
\begin{equation}
\ca U_{1} c_{j\downarrow} \ca U_{1}^{\dagger} = c_{j\downarrow}^{\dagger} \: , \qquad
\ca U_{1} c^{\dagger}_{j\downarrow} \ca U_{1}^{\dagger} = c_{j\downarrow} \: ,
\end{equation}
$\ca U_{2}$ is a sign transformation (or ($\pi,\pi$)-momentum boost \cite{Schneider_et_al_2012}) applied to the spin-$\uparrow$ particles,
\begin{equation}
\ca U_{2} c_{j\uparrow} \ca U_{2}^{\dagger} = z_{j} c_{j\uparrow} \: , \qquad
\ca U_{2} c^{\dagger}_{j\uparrow} \ca U_{2}^{\dagger} = z_{j} c^\dagger_{j\uparrow} \: ,
\end{equation}
where the sign factor $z_{j}=\pm 1$ for the two different sublattices of the bipartite square lattice.
One easily verifies that $\ca U_{1} \ca U_{2} H \ca U_{2}^{\dagger} \ca U_{1}^{\dagger} = - H + \mbox{const.}$ (if $T^{H}_{jj}=0$, $T^{H}_{jk}=T^{H}_{kj}$, $V^{H}_{jk}=V^{H}_{kj}$) where the additive constant results in an irrelevant global phase in the time-evolution operator.  
Finally, $\ca U_{3}$ is the anti-unitary time reversal. 
The Hubbard Hamiltonian itself is time-reversal invariant but 
\begin{equation}
\ca U_{3} e^{-i   H (t-t_{0})} \ca U_{3}^{\dagger} = e^{i  H(t-t_{0})} \; . 
\end{equation}
Together with the sign change of $H$ under $\ca U_{1} \ca U_{2}$, this proves the invariance of the time-evolution operator under the combined transformation $\ca U$.

We are interested in the site-dependent observable $n_{j\uparrow} = c^{\dagger}_{j\uparrow} c_{j\uparrow}$. 
Obviously, 
\begin{equation}
\ca U n_{j\uparrow} \ca U^{\dagger} = n_{j\uparrow} \; , 
\end{equation}
and thus we have the dynamical symmetry
\begin{equation}
\begin{split}
&\ \langle \Psi'(t_{0}) | e^{i  H  (t-t_{0})} n_{j\uparrow} e^{-i  H
(t-t_{0})} | \Psi'(t_{0}) \rangle\\
= &\ \langle \Psi(t_{0}) | e^{i  H  (t-t_{0})} n_{j\uparrow} e^{-i  H
(t-t_{0})} | \Psi(t_{0}) \rangle
\end{split}
\end{equation}
for initial states related via $| \Psi'(t_{0}) \rangle = \mathcal{U} | \Psi(t_{0}) \rangle$ (see Eq.\ (\ref{eqn:sym}) in Appendix \ref{sec:dsym}).
 
We compare the time evolutions for two different initial states which are defined as the ground states of the Hubbard model $\ca B$ (see Eq.\ (\ref{eqn:hamb})) where (i) the hopping between the different plaquettes is switched off ($\ff V^{B}=0$) and where (ii) an external field is switched on. 
The ``spin-excitation state'' $| {\text{N\'eel}} , p \rangle$ is the plaquette ground state in the presence of a staggered magnetic field:
\begin{equation}
\ca B \mapsto \ca B - h \sum_{j} z_{j} (n_{j\uparrow} - n_{{j\downarrow}}) \; ,
\label{eqn:spinexc}
\end{equation}
where $z_{j}=(-1)^{j}$.
The ``charge-excitation state'' $| {\rm CDW} ,p\rangle$ is obtained with the help of a staggered potential:
\begin{equation}
\ca B \mapsto \ca B - h \sum_{j} z_{j} (n_{j\uparrow} + n_{{j\downarrow}}) \; .
\label{eqn:chargeexc}
\end{equation}
The final-state dynamics for $t>t_{0}$ is governed by the Hubbard Hamiltonian $H$ with the hopping between the different plaquettes switched on, i.e.\ $V^{H}_{jk} = V = T$ between nearest neighbors $j$ and $k$, but with the field switched off, i.e.\ $h=0$.

\begin{figure}[t]
\includegraphics[width=0.43\textwidth]{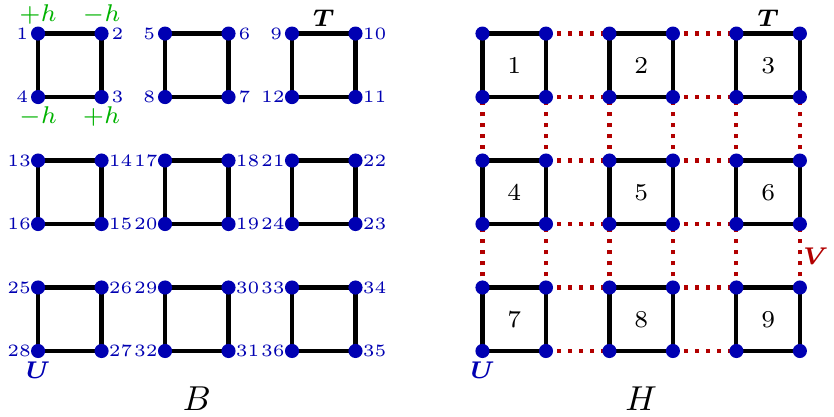}
\caption{\label{fig:system} (Color online) Pictorial representation of the Hamiltonians $B$ and $H$. The system is prepared as the ground state of $\ca B$ in the presence of a staggered field $h$ coupling to the $z$-component of the local spin or to the charge density. The evolution of the respective final states is given by the Hamiltonian $H$ with inter-plaquette hopping $V=T$ (nearest-neighbor hopping $T=-1$). Here, the decomposition into plaquettes is artificial and defines the NE-CPT scheme. Lattice sites as well as plaquettes are numbered. Interactions and hoppings are indicated exemplarily only.
}
\end{figure}

The actual NE-CPT calculations are performed for a finite system with $6\times 6$ sites and open boundary conditions.
For this system size the numerical computations can be performed conveniently on a standard desktop computer.
While the initial states are taken as simple product states (Fig.\ \ref{fig:system}, left) and are easily computed by means of exact diagonalization, an approximation must be used to determine the final-state dynamics:  
The inter-plaquette hopping $V=T$ is treated perturbatively to all orders within the CPT, see Fig.\ \ref{fig:system}, right.
One plaquette at a time, in the order indicated by the numbers, is coupled in a sequence of 8 NE-CPT steps.
We have checked that the results do not change within numerical accuracy when using a coupling scheme with a different ordering.
 
As $\ff V^{B}=0$ for the initial state, there are no potential-scattering vertices on the Matsubara branch of the contour.
Consequently, only the two Keldysh branches must be taken into account in the CPT equations, Eqs.\ (\ref{eqn:cptblock}).
For the numerical solution of the corresponding integral equation, we introduce a time discretization with a finite time step $\Delta t$. 
Eqs.\ (\ref{eqn:cptblock}) are considered as linear systems of equations for the different components of the CPT propagators. 
This component decomposition is advantageous as it allows an efficient use of standard quadrature rules.
Details are discussed in the Appendix \ref{sec:comp}.
Applying the trapezoidal rule, a time step $\Delta t = 0.05/|T|$ has turned out as sufficient for converged results.

\section{Numerical results and discussion}

\begin{figure}[b]
\includegraphics[width=0.475\textwidth]{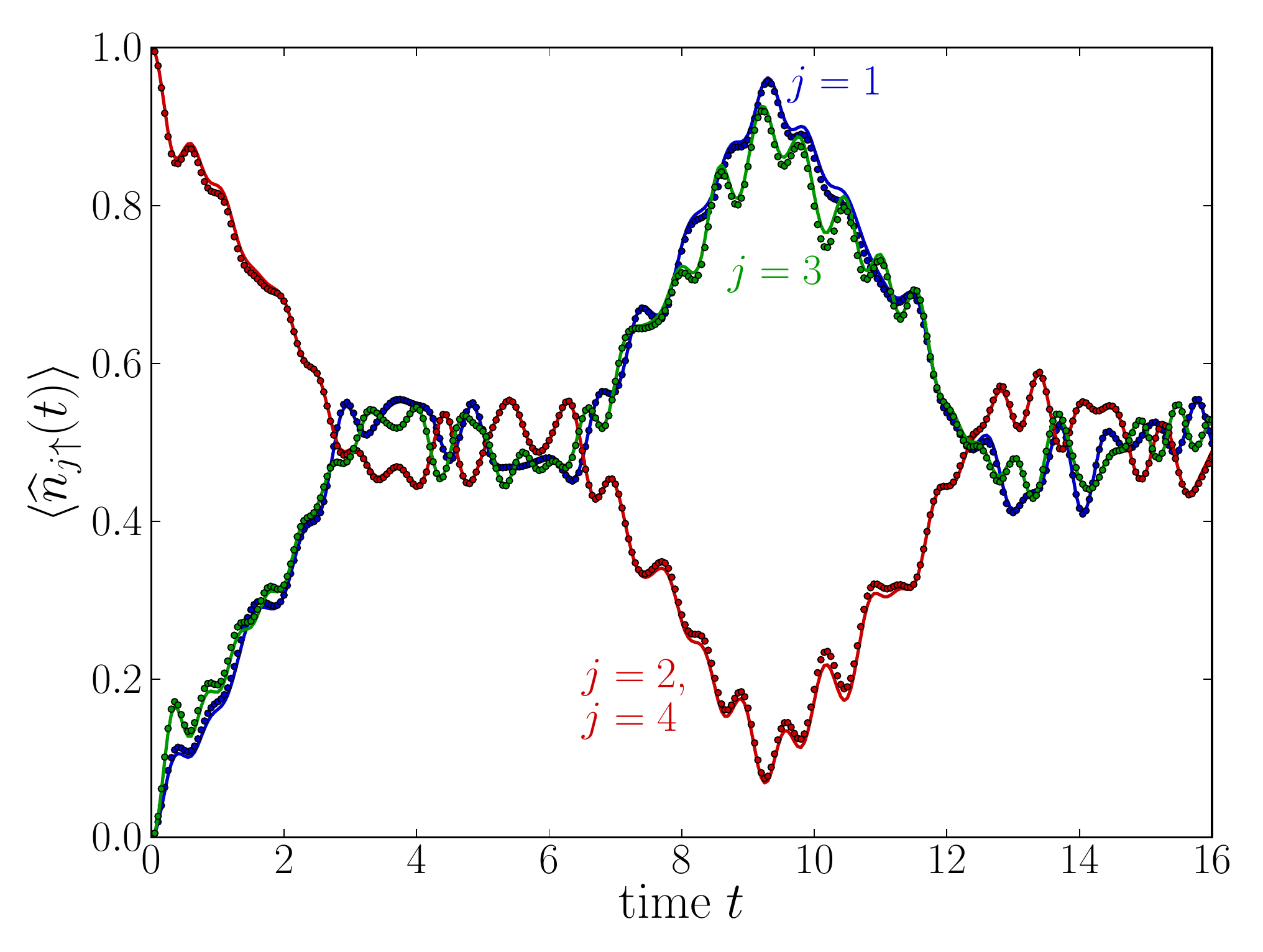}
\caption{\label{fig:results_glob} (Color online) 
Time evolution of the local spin-$\uparrow$ occupation numbers at the sites $j=1$, ..., $j=4$ for the system displayed in Fig.\ \ref{fig:system} (with the staggered field applied to every plaquette).
Inter-cluster hopping $V = T$, Hubbard interaction $U = 8|T|$. 
Energy and time scales are set by $T=-1$. 
Lines: initial ``spin-excitation state'' $|\text{N\'eel}\rangle$ obtained with $h=100$ in Eq.\ (\ref{eqn:spinexc}).
Dots: initial ``charge-excitation state'' $|\rm CDW\rangle$, $h=100$, see Eq.\ (\ref{eqn:chargeexc}).
}
\end{figure}

\begin{figure*}[t]
\includegraphics[width=0.95\textwidth]{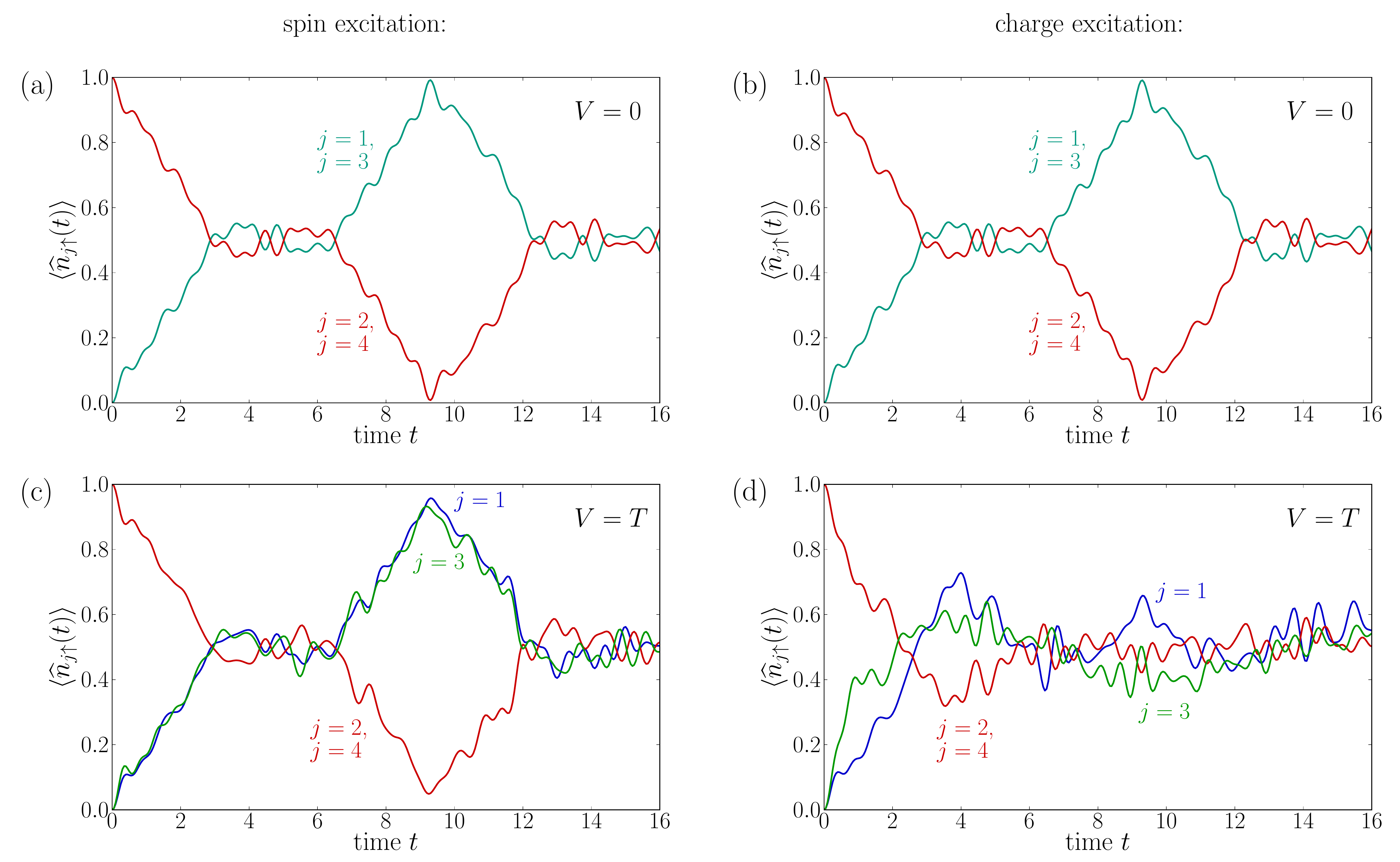}
\caption{\label{fig:results_loc} 
(Color online) 
Time evolution of the local spin-$\uparrow$ occupation numbers at the sites $j=1$, ..., $j=4$ as in Fig.\ \ref{fig:results_glob}  ($U = 8|T|$, $T=-1$)
but with the initial field (strength $h=100$) applied to the sites in plaquette $p=1$ only. 
The initial states of plaquettes $p=2$, ..., $p=9$ are the respective ground states $| \text{GS}, p \rangle$.
Top (a,b): Intra-plaquette dynamics only ($V=0$).
Bottom (c,d): Plaquettes coupled via NE-CPT ($V=T$).
Left (a,c): initial ``spin-excitation state'' on $p=1$, i.e.\ $|\text{N\'eel}, 1 \rangle$.
Right (b,d): initial ``charge-excitation state'' on $p=1$, i.e.\ $|\rm CDW, 1\rangle$.
}
\end{figure*}

We first consider a global excitation, i.e.\ $h$ is finite on all sites for the initial state, and the $j$-sums in Eqs.\ (\ref{eqn:spinexc}) and (\ref{eqn:chargeexc}) extend over the entire lattice.
In case of a small field strength $h$, we find clearly different time evolutions of the local spin-$\uparrow$ occupation number $\langle \widehat{n}_{j\uparrow} (t) \rangle$ for the two different initial states $|\text{N\'eel}\rangle = \otimes_{p} |\text{N\'eel}, p\rangle$ and $|\text{CDW}\rangle=\otimes_{p}|\text{CDW}, p\rangle$. 
In fact, the dynamical symmetry is expected to hold for $h \to \infty$ only. 
In this limit, a fully polarized N\'eel state $|\downarrow ,\uparrow ,\downarrow ,\uparrow \rangle$ and a CDW state
$| 0, \uparrow \downarrow ,0, \uparrow \downarrow \rangle$ are created on each plaquette, respectively, which are mapped onto each other:
$| \downarrow, \uparrow ,\downarrow ,\uparrow \rangle = \ca U | 0, \uparrow \downarrow ,0, \uparrow \downarrow \rangle$.

The results for a large field $h=100$ are shown in Fig.\ \ref{fig:results_glob}.
At $t=t_{0}=0$ we find 
$\langle \widehat{n}_{j\uparrow} (t) \rangle \approx 0$ for $j=1$ and $j=3$ and
$\langle \widehat{n}_{j\uparrow} (t) \rangle \approx 1$ for $j=2$ and $j=4$ 
(see Fig.\ \ref{fig:system} for the labeling of the sites) 
for both, the spin- and the charge-excitation state.
Note that the occupation numbers of the spin-$\downarrow$ particles are fixed by particle-hole and spin-inversion symmetry, i.e.\
$\langle \widehat{n}_{j\downarrow}(t) \rangle = 1 - \langle \widehat{n}_{j\uparrow}(t) \rangle$ (N\'eel) and $\langle \widehat{n}_{j\downarrow}(t) \rangle = \langle \widehat{n}_{j\uparrow}(t) \rangle$ (CDW).
Hence, total particle-number conservation and, in the same way, conservation of the $z$-component of the total spin is enforced. 
This symmetry has also been verified numerically. 
Furthermore, spatial symmetries (see Fig.\ \ref{fig:system}) enforce equal occupations $\langle \widehat{n}_{j\uparrow} (t) \rangle$ for the sites $j=2$ and $j=4$ while sites $j=1$ and $j=3$ are inequivalent. 
As can be seen from Fig.\ \ref{fig:results_glob}, this is respected by the NE-CPT. 
The approximate approach is also seen to respect the dynamical symmetry:
The results obtained for the two different initial states $|\text{N\'eel}\rangle$ and $|\rm CDW\rangle$, displayed as lines and dots in Fig.\ \ref{fig:results_glob}, respectively, are almost equal in the entire time interval studied.
Remaining discrepancies are small and are due to residual numerical errors in the solution of the CPT equation as well as due to tiny deviations from full polarization in the initial state at $h=100$.

Fig.\ \ref{fig:results_loc} (a) and (b) displays the time evolution of $\langle \widehat{n}_{j\uparrow} (t) \rangle$ starting from initial N\'eel and CDW states ($h=100$) for the sites in the isolated plaquette $p=1$: 
The inter-plaquette hopping stays switched off. 
Due to the dynamical symmetry, the time evolution is the same in both cases (a) and (b). 
In Fig.\ \ref{fig:results_loc} (c) and (d) the corresponding results are shown for an initial N\'eel and CDW state prepared on plaquette $p=1$ only ($h=100$) while plaquettes $p=2$, ..., $p=9$ are assumed to be in their singlet ground state $|{\rm GS},p\rangle$, i.e.\ the $j$-sums in Eqs.\ (\ref{eqn:spinexc}) and (\ref{eqn:chargeexc}) extend over $p=1$ only.
Contrary to (a) and (b), the inter-plaquette hopping is switched on ($V=T$) for times $t>0$.

We first note that $\ca U$ transforms the initial states on plaquette $p=1$ into each other, $|\text{N\'eel} , 1\rangle = \ca U |\rm CDW , 1 \rangle$, but does not leave the singlet ground state on the other plaquettes invariant, i.e.\ $|{\rm GS},p\rangle \ne \ca U | {\rm GS},p\rangle$. 
Consequently, the dynamical symmetry for the two different initial states is lost and different time evolutions are expected for finite $V$. 
This is nicely seen in Fig.\ \ref{fig:results_loc} by comparing (c) with (d).
It is striking, however, that in case of the initial N\'eel state (c) the time evolution of $\langle \widehat{n}_{j\uparrow} (t) \rangle$ for sites $j$ in plaquette $p=1$ does not deviate much from the pure intra-plaquette dynamics shown in (a). 
On the other hand, starting from the initial CDW state (d), we find strong inter-plaquette effects dominating the dynamics almost immediately after the quench.

For an explanation of these findings, we first concentrate on the intra-plaquette dynamics. 
Here, it is sufficient to discuss the case of the N\'eel initial state only:
As is seen in the figure, the full spin polarization quickly decays but at $t_{\rm r}(U) / 2 \approx 9.3$ for $U=8$ an almost fully polarized state with a reversed sign of the local spin is recovered. 
Note that $t_{\rm r}(U)$ is different from and actually much smaller than the exact recurrence time of the finite quantum system.
However, it does represent the exact recurrence time in the strong-coupling limit $U \to \infty$ where the low-energy sector of the plaquette Hamiltonian is exactly mapped onto an antiferromagnetic Heisenberg plaquette.

Consider a dimer model rather than a plaquette for a moment:
For $U \to \infty$, the N\'eel initial state $| \uparrow , \downarrow \rangle$ couples to the singlet ground state and to the triplet excited state of the effective Heisenberg model $H_{\rm low} = J \ff S_{1} \ff S_{2}$, namely 
$| \uparrow , \downarrow \rangle = \sqrt{1/2} ( | S=0, M=0 \rangle + | S=1, M=0 \rangle)$.
The time evolution of $\langle \widehat{n}_{j\uparrow} (t) \rangle = 0.5 + \langle \widehat{S}_{j,z} (t) \rangle$ is thus governed by a single frequency $\omega$ given by the excitation energy $\omega = E(S=1) - E(S=0) = J = 4 T^{2} / U$ which translates into a recurrence time $t_{\rm r}(U) = 2 \pi / \omega = \pi U / (2 T^{2})$.

For the plaquette $p=1$, $|\text{N\'eel} ,1\rangle$ turns out to be given by a linear superposition of three energy eigenstates for $U\to \infty$.
The resulting three excitation frequencies $\omega_{1,2,3}$, however, are integer multiples of the smallest one, say $\omega_{1}$. 
Rescaling the corresponding recurrence time $t_{\rm r}(U) = 2 \pi / \omega_{1} \propto U$ gives us $t_{r}(U=8) \approx 12.6$ which is close but not equal to the observed $t_{\rm r}(U) \approx 18.6$.
This is an effect of residual charge fluctuations which are absent in the pure spin model and small but non-negligible for $U=8$.
Charge fluctuations are also responsible for the small wiggles which can be seen in Fig.\ \ref{fig:results_loc} (a) on a time scale $2 \pi / U\lesssim 1$ and which are absent in the Heisenberg limit.

Having understood the physical origin of the different structures in case of isolated plaquettes, we can interpret the results in Fig.\ \ref{fig:results_loc} (c) and (d).
The initial N\'eel state on $p=1$ has a mean energy $\langle \text{N\'eel}, 1 | H | \text{N\'eel} ,1\rangle \sim J$ and thus preferably couples to the low-energy eigenstates of the full model while for the initial CDW state we have $\langle \text{CDW} ,1| H | \text{CDW}, 1 \rangle \sim 2U$, i.e.\ the latter mainly couples to highly excited states. 
Note that the energy spread $\Delta E = \sqrt{\langle H^{2} \rangle - \langle H \rangle^{2}} \approx 2.8$ is almost the same for both, the N\'eel and the CDW initial state. 

For the N\'eel initial state this implies that the overall trend of $\langle \widehat{n}_{j\uparrow} (t) \rangle$ is much slower as compared to the CDW initial state.
Therefore, on the time interval displayed in panel (c), the dynamics still follows the intra-plaquette dynamics (a) which is governed by the time scale $2\pi /J$.
The much less pronounced structures on the time scale $2\pi / U$ that are induced by residual charge fluctuations, on the other hand, are almost immediately affected by the scattering at the inter-plaquette potential $\ff V$. 
This is clearly visible, e.g.\ by comparing the results for inequivalent sites $j=1$ and $j=3$ in (c). 
Relaxation of $\langle \widehat{n}_{j\uparrow} (t) \rangle$ to its equilibrium value  $\langle {n}_{j\uparrow} \rangle = 0.5$ is thus expected for times much larger than $2\pi / J$. 

In case of the initial CDW state, the overall features of the intra-plaquette dynamics (b) are dominated by the same time scale $2\pi / J$ due to the dynamical symmetry.
For coupled plaquettes (d) where the dynamical symmetry is absent, however, the dominant time scale is rather given by $2\pi / U \ll 2\pi /J$. 
This explains the observed much faster relaxation of $\langle \widehat{n}_{j\uparrow} (t) \rangle$ on a time scale larger than $2\pi / U$ but smaller than $2\pi /J$.

We conclude that the largely different time evolutions displayed in (c) and (d) result from a clear separation of time scales while the dynamical symmetry enforces equal behavior for decoupled plaquettes as seen in (a) and (b). 
This interpretation is corroborated by calculations at a much weaker interaction strength $U=2$ (not shown) where nominally $J=U$ but where there are actually no well-defined local moments. 
Here, the relaxation of $\langle \widehat{n}_{j\uparrow} (t) \rangle$ is indeed equally fast for both types of initial states.
We also note that in case of a global excitation (see Fig.\ \ref{fig:results_glob}), where the dynamical symmetry is intact, the results for $\langle \widehat{n}_{j\uparrow} (t) \rangle$ again closely resemble those obtained for the isolated plaquette, see Fig.\ \ref{fig:results_loc} (a) and (b).

Finally, our cluster mean-field results for the two-dimensional Hubbard model can qualitatively be compared with numerically exact data obtained by means of time-dependent density-matrix renormalization-group methods (DMRG) for the one-dimensional Hubbard model. \cite{dmrg1,dmrg2}
In Ref.\ \onlinecite{dmrg1} a local doublon-holon excitation of the ground state of the half-filled model has been considered. 
The doublon is found to delocalize quickly on a time scale of a few $1/|T|$, very similar to doublon delocalization in an empty-band system, \cite{HP12} but, on a scale of several tens of $1/|T|$, does not completely {\em decay} as higher-order many-body scattering processes are necessary in the strong-coupling regime to break up the repulsively bound fermion pair consistent with energy conservation.
Note that within the NE-CPT there is no direct access to two-particle expectation values.
Nevertheless, the fast delocalization of the initial charge excitation is qualitatively consistent with our results in Fig.\ \ref{fig:results_loc} (d). 

For the homogeneously excited initial state $| \rm CDW \rangle$ where a doublon is present at every second lattice site, DMRG \cite{dmrg2} again predicts the doublons as very stable in the strong-coupling regime, i.e.\ the total double occupation per site $\langle \widehat{D}(t) \rangle$ relaxes to a constant which is close to its maximum initial value $\langle \widehat{D}(0) \rangle = 0.5$ (an exponential decay is only found in the presence of a finite nearest-neighbor Coulomb interaction). 
Our results (Fig.\ \ref{fig:results_glob}) are again qualitatively consistent with the DMRG data:
At $t_{1} \approx 9.3$ we find the local occupation numbers close to unity ($j=1,3$) or zero ($j=2,4$). 
Here, we can easily estimate $\langle \widehat{D}(t_{1}) \rangle \approx \sum_{j\in p} \langle \widehat{n}_{j\uparrow} (t_{1}) \rangle \langle \widehat{n}_{j\downarrow} (t_{1}) \rangle / L_{p} \lesssim 0.5$.
As the occupation-number dynamics $\langle \widehat{n}_{j\sigma} (t) \rangle$ does not differ very much from the dynamics for an isolated plaquette on the time interval considered here, we expect the same for $\langle \widehat{D}(t) \rangle$.
For the isolated plaquette, we in fact find $\langle \widehat{D}(t) \rangle$ slightly oscillating around a value smaller but close to 0.5.

\section{Conclusions}

The generalization of the cluster-perturbation theory to systems with a non-trivial real-time dynamics of an initial non-equilibrium state provides a conceptually appealing approach that has been demonstrated as numerically feasible even for two-dimensional lattice models of strongly correlated fermions.
The NE-CPT represents a non-self-consistent cluster mean-field-type approach where the effect of the scattering at the inter-cluster potential on the self-energy is neglected. 
For the particle-hole and spin symmetric Hubbard model, it nevertheless respects macroscopic conservation laws and has previously been shown \cite{Balzer_et_al_2012} to be able to reliably describe relaxation dynamics on short time scales for one-dimensional and impurity-type models.

The present study has addressed important numerical improvements that are necessary to study two-dimensional models: 
First, for a numerically efficient evaluation of the NE-CPT equations on the Keldysh-Matsubara contour, we have rewritten and implemented the NE-CPT equations for the different components of the non-equilibrium Green's function. 
This leads to a considerable increase of the numerical accuracy and the accessible times. 
Secondly and more important, however, we have demonstrated that standard concepts of multiple-scattering theory, known from {\em ab initio} electronic-structure theory, can successfully be applied to the non-equilibrium case. 
The iterative scheme of coupling of only two equal or different building blocks at a time represents an efficient and flexible concept that is applicable to the two-dimensional case but may also be useful in different and even more complicated situations, e.g.\ in cases with fewer spatial symmetries.

As a concrete example, the dynamics of the Hubbard model on a $6\times 6$ square array has been considered for initial states prepared as plaquette-product states.
We have studied the time evolution of the spin-dependent local occupation numbers to understand relaxation effects induced by inter-plaquette correlations building up in the course of time.
Their real-time dynamics in the two states $|\text{N\'eel}\rangle = \otimes_{p} |\text{N\'eel}, p\rangle$ and $|\text{CDW}\rangle=\otimes_{p}|\text{CDW}, p\rangle$ with a {\em global} staggered spin or charge excitation is found to be identical. 
This is enforced by a dynamical symmetry, i.e.\ an invariance of the time-evolution operator and of the considered observables under a combined spin-asymmetric particle-hole, staggered sign, and time-reversal transformation $\ca U$, which is found to be respected by the NE-CPT.
This dynamical symmetry prevents a fast relaxation of the CDW initial state which was expected to take place on a short time scale $2\pi / U$ related to charge dynamics. 

If, on the other hand, the dynamical symmetry is broken by preparing the spin or charge excitation {\em locally} on the plaquette $p=1$ only while the rest of the system is given as a product of plaquette ground states $|\text{GS}, p\rangle$, a clear separation of energy and time scales is observed in the relaxation dynamics: 
The occupation numbers in the spin-excited plaquette ($|\text{N\'eel}, p=1\rangle$) show an overall slow coupling to the environment on the scale $2\pi/J$, with some fast but small oscillations due to residual charge fluctuations on top.
Contrary, a charge dynamics on the scale $2\pi/U \ll 2\pi /J$ is seen to result in a fast relaxation of the of occupation numbers for the charge-excited state $|\text{CDW}, p=1\rangle$.

One might speculate on the long-time limit that is not accessible to real-time Green's-function-based approaches as the NE-CPT:
For the global spin and charge excited states, the dynamical symmetry enforces equal time evolutions although the total energies per site are largely different ($\sim U / 2$) which therefore implies different thermal states characterized by different temperatures eventually.
In fact, the dynamical symmetry translates into a symmetry of the thermal states as $H\mapsto -H + {\rm const.}$ under $\ca U$ implies a transformation of the (normalized) thermal mixed state $\rho \mapsto \rho'$ with a negative temperature $T \mapsto -T$. 
For the locally excited states, on the other hand, the higher energy of the local charge excitation is expected to dissipate to the bulk of the system resulting, in the thermodynamical limit, in equal temperatures and equal thermal states eventually. 

\begin{acknowledgments}
We would like to thank M. Balzer and H. Moritz for instructive discussions.
The work was supported by the Deutsche Forschungsgemeinschaft within the Sonderforschungsbereich 925 (project B5) and within the excellence cluster ``The Hamburg Centre for Ultrafast Imaging - Structure, Dynamics and Control of Matter at the Atomic Scale''.
\end{acknowledgments}

\appendix

\section{Dynamical symmetries}
\label{sec:dsym}

A unitary or anti-unitary transformation $\ca U$ of the observables and of the states of a quantum system leaves measurable quantities, such as expectation values, invariant. \cite{Messiah_1961,Mertsching_1977}
The quantum system itself is said to be ``symmetric'' with respect to the transformation, if its Hamiltonian $H$ is invariant: $H'=\ca U H \ca U^{\dagger} = H$.
For continuous transformation groups with $\ca U = \exp(i\ff \Lambda\ff \varphi)$ and para\-meters $\ff \varphi$, the invariance implies that the corresponding generators $\ff \Lambda=\ff \Lambda^{\dagger}$ commute with $H$, i.e.\ $[\ff \Lambda, H] = 0$. 
If the Hamiltonian is not explicitly time-dependent, this leads to conservation laws of the form 
$\langle \Psi(t) | \ff \Lambda | \Psi(t) \rangle = \langle \Psi(t_{0}) | \ff \Lambda | \Psi(t_{0}) \rangle = \mbox{const.}$, where $| \Psi(t) \rangle = \exp(-i  H  (t-t_{0})) | \Psi(t_{0}) \rangle$.

This concept might be generalized to ``dynamical symmetries'' in the following way:
Consider a transformation $\ca U$ that leaves an observable invariant, 
\begin{equation}
A \to A' =\ca U A \ca U^{\dagger} = A \; , 
\end{equation}
as well as the time-evolution operator:
\begin{equation}
\begin{split}
e^{-i  H  (t-t_{0})} \to (e^{-i  H (t-t_{0})})' &= \ca U e^{-i  H  (t-t_{0})} \ca U^{\dagger}\\
&= e^{-i  H  (t-t_{0})} \; .
\end{split}
\label{eqn:timeevo}
\end{equation}
This immediately implies that
\begin{equation}
\begin{split}
&\ \langle \Psi'(t_{0}) | e^{i  H  (t-t_{0})} A e^{-i  H  (t-t_{0})} | \Psi'(t_{0}) \rangle\\
= &\ \langle \Psi(t_{0}) | e^{i  H  (t-t_{0})} A e^{-i  H  (t-t_{0})} | \Psi(t_{0}) \rangle \quad ,
\label{eqn:sym}
\end{split}
\end{equation}
for $t \geq t_{0}$, i.e.\ the time evolution is the same for different initial states,
$|\Psi'(t_{0}) \rangle$ and $| \Psi(t_{0}) \rangle$.

\section{Component decomposition}
\label{sec:comp}

The $\bullet$~operation occurring in the CPT equation (Eq.~(\ref{eqn:cpt-eqn})) includes a convolution with respect to time variables.
The corresponding integration path is the contour $\ca C$ shown in Fig.~\ref{fig:contour}. 
If $\ca C$ was considered as a whole, the error due to the time discretization in the numerical treatment of the integrals, e.g.\ via a Newton-Cotes formula, would be dominated by the characteristic discontinuities of non-equilibrium Green's functions originating from the time ordering along $\ca C$, see Eq.\ (\ref{eqn:gfdef}).
Smaller errors or larger time discretization steps and thus a larger $t_{\text{max}}$, however, are attainable by splitting the non-equilibrium Green's functions into certain components based on the different branches of $\mathcal{C}$. 
Exemplified by $\ff G$ and omitting spin and spatial indices for simplicity, one may define
\begin{subequations}
\begin{align}
G^{>}(t_{1},t_{2}) &\equiv G(t_{1}^{-},t_{2}^{+})\; , \\
G^{<}(t_{1},t_{2}) &\equiv G(t_{1}^{+},t_{2}^{-})\; , \\
G^{\urcorner}(t_{1},\tau_{2}) &\equiv G(t_{1}^{\pm},t_{0}\!\!-\!i\tau_{2})\; , \\
G^{\ulcorner}(\tau_{1},t_{2}) &\equiv G(t_{0}\!\!-\!i\tau_{1},t_{2}^{\pm})\; , \\
G^{\text{\tiny MG}}(\tau_{1},\tau_{2}) &\equiv -i\,\left\langle \widehat{c}\,(t_{0}\!\!-\!i\tau_{1}) \ \widehat{c}^{\dagger}(t_{0}\!\!-\!i\tau_{2}) \right\rangle\; , \\
G^{\text{\tiny ML}}(\tau_{1},\tau_{2}) &\equiv i\,\left\langle \widehat{c}^{\dagger}(t_{0}\!\!-\!i\tau_{2}) \ \widehat{c}\,(t_{0}\!\!-\!i\tau_{1}) \right\rangle\; , \\
G^{\text{ret}}(t_{1},t_{2}) &\equiv \Theta(t_{1}\!\!-\!t_{2}) \left(G^{>}(t_{1},t_{2})\!-\!G^{<}(t_{1},t_{2})\right) \; , \\
G^{\text{adv}}(t_{1},t_{2}) &\equiv -\Theta(t_{2}\!\!-\!t_{1}) \left(G^{>}(t_{1},t_{2})\!-\!G^{<}(t_{1},t_{2})\right) \: .
\end{align}
\end{subequations}
Besides the distinction between ``greater'' and ``lesser'' Matsubara components $G^{\text{\tiny MG}}$ and $G^{\text{\tiny ML}}$, the component definitions are well-known in non-equilibrium Green's function literature. \cite{vanLeeuwen_et_al_2005} 
For our present purposes, $G^{\text{ret}}$ and $G^{\text{adv}}$ are auxiliary quantities only,  
all necessary information on $\ff G$ is already contained in the components $G^{>}$, ..., $G^{\text{\tiny ML}}$. 
We furthermore write 
\begin{equation}
\bigl(A^{c_{1}} \circ_{t_{3}}^{t_{4}} B^{c_{2}}\bigr)(z_{1},z_{2}) 
\equiv 
\int_{t_{3}}^{t_{4}}\!dt_{5} A^{c_{1}}(z_{1},t_{5})B^{c_{2}}(t_{5},z_{2}) \: ,
\end{equation}
and
\begin{equation}
\bigl(A^{c_{1}} \ast_{\tau_{3}}^{\tau_{4}} B^{c_{2}}\bigr)(z_{1},z_{2}) \equiv -i\int_{\tau_{3}}^{\tau_{4}}\!d\tau_{5} A^{c_{1}}(z_{1},\tau_{5})B^{c_{2}}(\tau_{5},z_{2}) \: .
\end{equation}
With this, a contour convolution $C(z_{1},z_{2}) \equiv \int_\mathcal{C}dz_{3} \, A(z_{1},z_{3}) \, B(z_{3},z_{2})$ can be split into the different components and reads:
\begin{subequations}
\begin{align}
\label{eqn:conv-gre}&C^{>} =\, A^{>} \underset{t_{0}}{\overset{t_{\text{max}}}{\circ}} B^{\text{adv}} + A^{\text{ret}} \underset{t_{0}}{\overset{t_{\text{max}}}{\circ}} B^{>} + A^{\urcorner} \underset{0}{\overset{\beta}{\ast}} B^{\ulcorner} \, ,\\[-2pt]
\label{eqn:conv-les}&C^{<} =\, A^{<} \underset{t_{0}}{\overset{t_{\text{max}}}{\circ}} B^{\text{adv}} + A^{\text{ret}} \underset{t_{0}}{\overset{t_{\text{max}}}{\circ}} B^{<} + A^{\urcorner} \underset{0}{\overset{\beta}{\ast}} B^{\ulcorner} \, ,\\[-2pt]
\label{eqn:conv-hoole}&C^{\urcorner} =\, A^{\text{ret}} \underset{t_{0}}{\overset{t_{\text{max}}}{\circ}} B^{\urcorner} + A^{\urcorner} \underset{0}{\overset{\tau_{2}}{\ast}} B^{\text{\tiny ML}} + A^{\urcorner} \underset{\tau_{2}}{\overset{\beta}{\ast}} B^{\text{\tiny MG}} \, ,\\[-2pt]
\label{eqn:conv-hoori}&C^{\ulcorner} =\, A^{\ulcorner} \underset{t_{0}}{\overset{t_{\text{max}}}{\circ}} B^{\text{adv}} + A^{\text{\tiny MG}} \underset{0}{\overset{\tau_{1}}{\ast}} B^{\ulcorner} + A^{\text{\tiny ML}} \underset{\tau_{1}}{\overset{\beta}{\ast}} B^{\ulcorner} \, ,\\[-2pt]
\label{eqn:conv-mgre}&C^{\text{\tiny MG}} \!\!\!\!\overset{\tau_{1} \underset{\downarrow}{>} \tau_{2}}{=}\!\!\! A^{\text{\tiny MG}} \underset{0}{\overset{\tau_{2}}{\ast}} B^{\text{\tiny ML}} + A^{\text{\tiny MG}} \underset{\tau_{2}}{\overset{\tau_{1}}{\ast}} B^{\text{\tiny MG}} + A^{\text{\tiny ML}} \underset{\tau_{1}}{\overset{\beta}{\ast}} B^{\text{\tiny MG}} \, ,\\[-2pt]
\label{eqn:conv-mles}&C^{\text{\tiny ML}} \!\!\!\!\overset{\tau_{1} \underset{\downarrow}{\leq} \tau_{2}}{=}\!\!\! A^{\text{\tiny MG}} \underset{0}{\overset{\tau_{1}}{\ast}} B^{\text{\tiny ML}} + A^{\text{\tiny ML}} \underset{\tau_{1}}{\overset{\tau_{2}}{\ast}} B^{\text{\tiny ML}} + A^{\text{\tiny ML}} \underset{\tau_{2}}{\overset{\beta}{\ast}} B^{\text{\tiny MG}} \, ,\\[-2pt]
\label{eqn:conv-ret}&C^{\text{ret}} =\, A^{\text{ret}} \underset{t_{0}}{\overset{t_{\text{max}}}{\circ}} B^{\text{ret}} \, ,\\[-2pt]
\label{eqn:conv-adv}&C^{\text{adv}} =\, A^{\text{adv}} \underset{t_{0}}{\overset{t_{\text{max}}}{\circ}} B^{\text{adv}} \, .
\end{align}
\end{subequations}
Here, we have also omitted time variables except for those appearing as integration boundaries.
$\tau_{\alpha}$ denotes the $\alpha^{\text{th}}$ argument of the component $C^{c}$. 

These convolution rules represent an adaption of the Langreth rules\cite{Langreth_and_Wilkins_1972} for our purposes and are similar to corresponding rules described in Ref.~\onlinecite{vanLeeuwen_et_al_2005}.
Application to the differential equation of motion for $\ff G$ yields generalized Kadanoff-Baym equations. \cite{Kadanoff_and_Baym_1962} 
Utilizing the component decomposition and also evaluating the unit step functions of the retarded and advanced components, we find all integrands being continuous functions of the respective time argument.
Furthermore, causality is made explicit by the (renewed) integration boundaries. 
Although the convolution $C$ does not behave like a proper Green's function in all aspects, it can play the role of $A$ or $B$ in a next convolution step, e.g. the components $C^{c}$ are continuous themselves. 

For the NE-CPT, we actually have to solve a problem of the form $\ff A \bullet \ff B = \ff C$ where $\ff A$ and $\ff C$ are given.
To this end, we rewrite the CPT equation for the different components and treat the resulting problem as a system of linear equations $(\ff 1 - \ff G' \bullet \ff V) \bullet \ff G^{\text{(CPT)}} = \ff G'$ which must be solved for $\ff G^{\text{(CPT)}}$. 
Because of the homogeneity of the Matsubara Green's function, solving Eq.~(\ref{eqn:conv-mgre}) with $\tau_{2} = 0$ is sufficient to get $B^{\text{\tiny MG}}$ and $B^{\text{\tiny ML}}$. 
These components in turn provide us with $B^{\urcorner}$ via Eq.~(\ref{eqn:conv-hoole}). 
Equipped with $B^{\text{adv}}$ from Eq.~(\ref{eqn:conv-adv}), Eq.~(\ref{eqn:conv-hoori}) can be solved for $B^{\ulcorner}$. 
Both components are prerequisites for finally obtaining $B^{>}$ and $B^{<}$ from Eqs.\ (\ref{eqn:conv-gre}) and (\ref{eqn:conv-les}), respectively. 
The lesser component is our main quantity of interest as it enters Eq.~(\ref{eqn:gf-obs}).
The mentioned solution steps can be performed separately for all spatial/temporal indices $k$ and $z_{2}$ of $B^{\sigma}_{jk}(z_{1},z_{2})$ with the additional benefit of recurring coefficient matrices. 
Furthermore, use can be made of the fact that coefficient matrices representing $A^{\text{ret}}$ and $A^{\text{adv}}$ are triangular.

\end{document}